# A Fuzzy Similarity Based Concept Mining Model for Text Classification

Text Document Categorization Based on Fuzzy Similarity Analyzer and Support Vector Machine Classifier

Shalini Puri
M. Tech. Student
BIT, Mesra
India

*Abstract*—**Text Classification is a challenging and a red hot field in the current scenario and has great importance in text categorization applications. A lot of research work has been done in this field but there is a need to categorize a collection of text documents into mutually exclusive categories by extracting the concepts or features using supervised learning paradigm and different classification algorithms. In this paper, a new Fuzzy Similarity Based Concept Mining Model (FSCMM) is proposed to classify a set of text documents into pre - defined Category Groups (CG) by providing them training and preparing on the sentence, document and integrated corpora levels along with feature reduction, ambiguity removal on each level to achieve high system performance. Fuzzy Feature Category Similarity Analyzer (FFCSA) is used to analyze each extracted feature of Integrated Corpora Feature Vector (ICFV) with the corresponding categories or classes. This model uses Support Vector Machine Classifier (SVMC) to classify correctly the training data patterns into two groups; i. e., + 1 and – 1, thereby producing accurate and correct results. The proposed model works efficiently and effectively with great performance and high - accuracy results.**

*Keywords-Text Classification; Natural Language Processing; Feature Extraction; Concept Mining; Fuzzy Similarity Analyzer; Dimensionality Reduction; Sentence Level; Document Level; Integrated Corpora Level Processing.*

I. INTRODUCTION

From the long time, the discipline of Artificial Intelligence (AI) is growing up on the map of science with psychology and computer science. It is an area of study that embeds the computational techniques and methodologies of intelligence, learning and knowledge [1] to perform complex tasks with great performance and high accuracy. This field is fascinating because of its complementarities of art and science. It contributes to increase the understanding of reasoning, learning and perception. Natural Language Processing (NLP) is the heart of AI and has text classification as an important problem area to process different textual data and documents by finding out their grammatical syntax and semantics and representing them in the fully structured form [1] [2]. AI provides many learning methods and paradigms to represent, interpret and acquire domain knowledge to further help other documents in learning.

Text Mining (TM) is a new, challenging and multi-disciplinary area, which includes spheres of knowledge like Computing, Statistics, Predictive, Linguistics and Cognitive Science [3] [4]. TM has been applied in a variety of concerns and applications. Some applications are summary creation, clustering, language identification [5], term extraction [6] and categorization [5] [6], electronic mail management, document management, and market research with an investigation [3] [4] [5].

TM consists of extracting regularities, patterns, and categorizing text in large volume of texts written in a natural language; therefore, NLP is used to process such text by segmenting it into its specific and constituent parts for further processing [2]. Text segmentation is also an important concern of TM. Many researches go on for the work of text classification [3] [4] [6] [7] [8] just for English language text. Text classification is performed on the textual document sets written in English language, one of the European Language, where words can be simply separated out using many delimiters like comma, space, full stop, etc. Most of the developed techniques work efficiently with European languages where the words, the unit of representation, can be clearly determined by simple tokenization techniques. Such text is referred as segmented text [10]. It does not always happen. There are some Asian Languages in which textual document does not follow word separation schemes and techniques. These languages contain un-segmented text. There are so many other languages in the world like Chinese and Thai Languages in which there is no delimiter to separate out the words. These languages are written as a sequence of characters without explicit word boundary delimiters [10]. So, they use different mechanisms for segmentation and categorization. Here, the proposed effort is to work only on the English text documents.

Therefore, TM categorization is used to analyze and comprise of large volume of non - structured textual documents. Its purpose is to identify the main concepts in a text document and classifying it to one or more pre-defined categories [3]-[12]. NLP plays an important and vital role to convert unstructured text [4] [5] [6] into the structured one by performing a number of text pre-processing steps. This processing results into the extraction of specific and exclusive concepts or features as words. These features help in categorizing text documents into classified groups.

Section II discusses the thematic background and related research work done on the concept mining, feature extraction, and similarity measure. In section III, the proposed model and methodology is discussed in detail. It discusses Text Learning Phase which includes Text Document Training Processor (TDTP), Pseudo Thesaurus, Class Feeder (CF) and Fuzzy Feature Category Similarity Analyzer (FFCSA). Next, Support Vector Machine Classifier (SVMC) and Text Classification Phase are discussed. Finally, section IV concludes the paper with the suggested future work and scope of the paper.

## II. THEMATIC BACKGROUND AND RELATED WORK

Supervised learning techniques and methodologies for automated text document categorization into known and predefined classes have received much attention in recent years. There are some reasons behind it. Firstly, in the unsupervised learning methods, the document collections have neither predefined classes nor labeled document's availability. Furthermore, there is a big motivation to uncover hidden category structure in large corpora. Therefore, text classification algorithms are booming up for word based representation of text documents and for text categorization.

### A. Concept Mining

Concept Mining is used to search or extract the concepts embedded in the text document. These concepts can be either words or phrases and are totally dependent on the semantic structure of the sentence. When a new text document is introduced to the system, the concept mining can detect a concept match from this document to all the previously processed documents in the data set by scanning the new document and extracting the matching concepts [5]. In this way, the similarity measure is used for concept analysis on the sentence, document, and corpus levels.

These concepts are originally extracted by the semantic role labeler [5] and analyzed with respect to the sentence, document, and corpus levels. Thus, the matching among these concepts is less likely to be found in non - related documents. If these concepts show matching in unrelated documents, then they produce errors in terms of noise. Therefore, when text document similarity is calculated, the concepts become insensitive to noise.

### B. Feature Extraction

In text classification, the dimensionality of the feature vector is usually huge. Even more, there is the problem of *Curse of Dimensionality*, in which the large collection of features takes very much dimension in terms of execution time and storage requirements. This is considered as one of the problems of *Vector Space Model (VSM)* where all the features are represented as a vector of *n* - dimensional data. Here, *n* represents the total number of features of the document. This features set is huge and high dimensional.

There are two popular methods for feature reduction: *Feature Selection and Feature Extraction*. In feature selection methods, a subset of the original feature set is obtained to make the new feature set, which is further used for the text classification tasks with the use of Information Gain [5]. In feature extraction methods, the original feature set is converted into a different reduced feature set by a projecting process. So, the number of features is reduced and overall system performance is improved [6].

Feature extraction approaches are more effective than feature selection techniques but are more computationally expensive. Therefore, development of scalable and efficient feature extraction algorithms is highly demanded to deal with high-dimensional document feature sets. Both feature reduction approaches are applied before document classification tasks are performed.

### C. Similarity Measure

In recent years, fuzzy logic [1] [2] [4] [6]-[15] has become an upcoming and demanding field of text classification. It has its strong base of calculating membership degree, fuzzy relations, fuzzy association, fuzzy production rules, fuzzy k-means, fuzzy c-means and many more concerns. As such, a great research work has been done on the fuzzy similarity and its classifiers for text categorization.

The categorizer based on fuzzy similarity methodology is used to create categories with a basis on the similarity of textual terms [4]. It improves the issues of linguistic ambiguities present in the classification of texts. So, it creates the categories through an analysis of the degree of similarity of the text documents that are to be classified. The similarity measure is used to match these documents with pre-defined categories [4] - [15]. Therefore, the document feature matrix is formed to check that a document satisfies how many defined features of the reduced feature set and categorized into which category or class [4] [6] [7]. The fuzzy similarity measure can be used to compute such different matrices.

## III. THE FUZZY SIMILARITY BASED CONCEPT MINING MODEL (FSCMM)

In this section, the proposed *Fuzzy Similarity Based Concept Mining Model (FSCMM)* is discussed. This model automatically classifies a set of known text documents into a set of category groups. The model shows that how these documents are trained step by step and classified by the Support Vector Machine Classifier (SVMC). SVMC is further used to classify various new and unknown text documents categorically.

The proposed model is divided into the two phases: *Text Learning Phase (TLP)* and *Text Classification Phase (TCP)*. TLP performs the learning function on a given set of text documents. It performs the steps of first stage; i. e., *Text Document Training Processor (TDTP)* and then the steps of second stage; i. e., *Fuzzy Feature Category Similarity Analyzer (FFCSA)*. The TDTP is used to process the text document by converting it into its small and constituent parts or chunks by using NLP concepts at the *Sentence, Document and Integrated Corpora Level*s. Then, it searches and stores the desired, important and non-redundant concepts by removing stop words, invalid words and extra words. In the next step, it performs word stemming and feature reduction. The result of sentence level preparation is low dimensional *Reduced Feature Vector (RFV)*. Each RFV of a document is sent for document level preparation, so that *Integrated Reduced Feature Vector (IRFV)* is obtained. To obtain IRFV, all the RFVs are integrated into one. Now, *Reduced Feature Frequency Calculator (RFFC)* is used to calculate the total frequency of each different word occurred in the document. Finally, all redundant entries of each exclusive word are removed and all the words with their associated frequencies are stored in decreasing order of their frequencies. At the integrated corpora level, the low dimension *Integrated Corpora Feature Vector (ICFV)* is resulted.

In such a way, feature vectors at each level are made low dimensional by processing and updating step by step. Such functionality helps a lot to search the appropriate concepts with reduced vector length to improve system performance and accuracy.

FFCSA performs similarity measure based analysis for feature pattern (TD – ICFV) using the enriched fuzzy logic base. The membership degree of each feature is associated with it. Therefore, an analysis is performed between every feature of a text document and class.

SVMC is used for the categorization of the text documents. It uses the concept of hyper planes to identify the suitable category. Furthermore, SVMC accuracy is checked by providing some new and unknown text documents to be classified into the respective Category Group (CG). This task is performed in TCP.

The proposed Fuzzy Similarity Based Concept Mining Model (FSCMM) is shown in Fig. 1. In the next sections, this model and its components are discussed in detail.

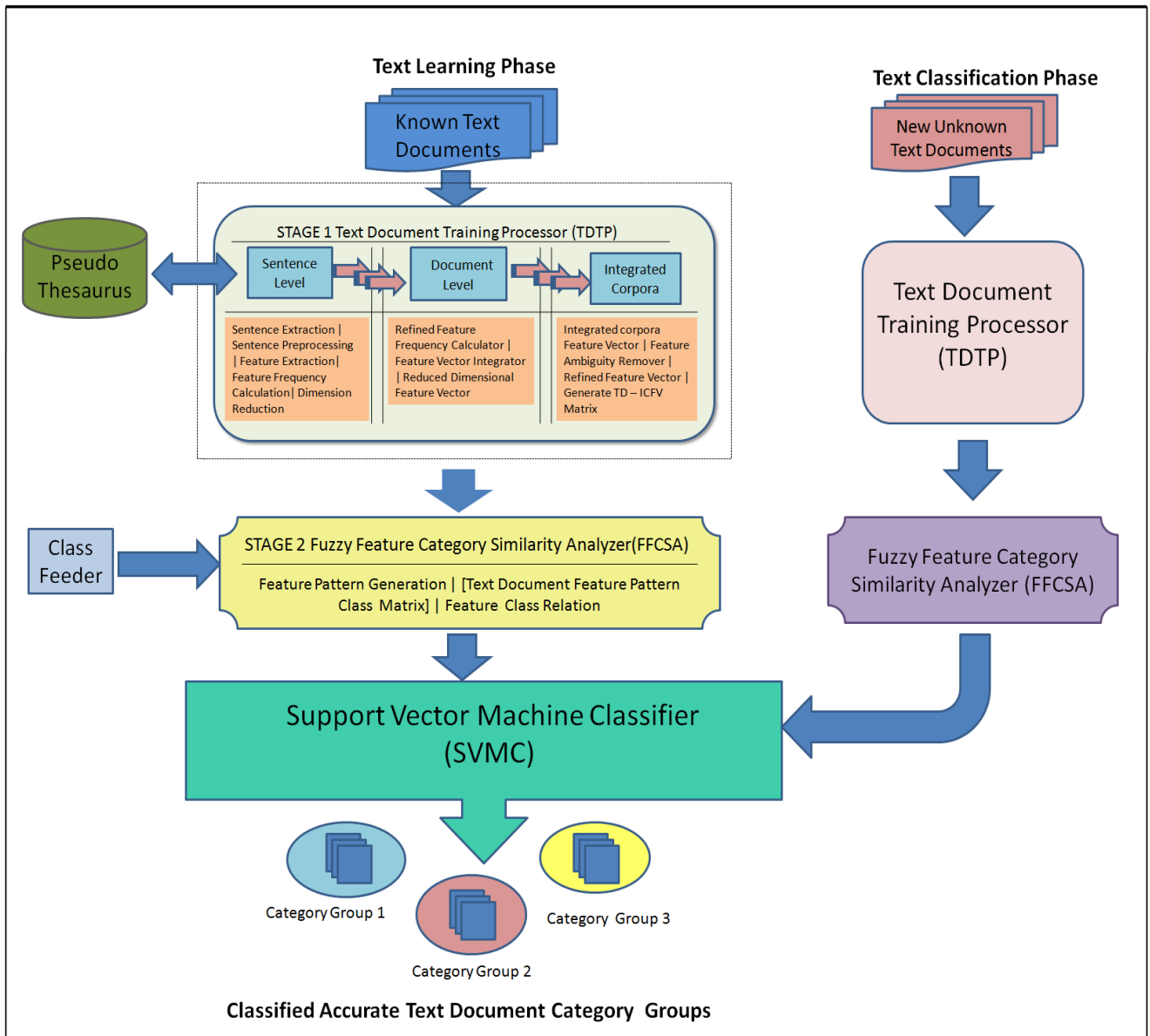

Figure 1. The Fuzzy Similarity Based Concept Mining Model (FSCMM).

## A. Text Learning Phase (TLP)

Consider a set of *n* text documents,

$$TD = \{TD_1, TD_2, TD_3, \ldots, TD_n\} \quad (1)$$

Where $TD_1, TD_2, TD_3, \ldots, TD_n$ are the individual and independent text documents which are processed, so that they can be categorized into the required category.

### 1) Text Document Training Processor (TDTP)

Text Document Training Processor (TDTP) prepares the given text document set *TD* of *n* text documents by performing many operations on the sentence, document, and integrated corpora levels. Firstly, each text document $TD_i$, $1 \leq i \leq n$, is processed at its sentence level. The result of such sentence level pre-processing for all the sentences of $TD_i$ is integrated into one, which is further processed and refined to make available for the integrated corpora. Integrated corpora accept and integrate all the refined text documents and perform more processing. Its result is sent to FFCSA.

#### a) At Sentence Level

This section describes that how a sentence is pre-processed and finds out the feature vector, each feature's frequency and finally, the conversion of the *Feature Vector (FV)* into the *Reduced Feature Vector (RFV)*.

A text document $TD_i$ is composed of a set of sentences, so consider

$$TD_i = \{s_{i1}, s_{i2}, s_{i3}, \ldots, s_{im}\} \quad (2)$$

Where *i* denotes the text document number and *m* denotes the total number of sentences in $TD_i$. *Sentence Extractor (SE)* is used to extract the sentence $s_{i1}$ from $TD_i$. Each sentence has its well-defined and non-overlapping boundaries, which makes the sentence extraction a simple task for SE.

When the sentence is extracted, a verb – argument structure is made for $s_{i1}$. The syntax tree is drawn using the pre-defined syntactic grammar to separate the verbs and the arguments of the sentence. The sentence can be composed of the nouns, proper nouns, prepositions, verbs, adverbs, adjectives, articles, numerals, punctuation and determiners. So, with the construction of the syntax tree, the stop word and other extra terms are removed except the nouns, proper nouns and numerals which are considered as the concepts. To remove the invalid and extra words, the *Pseudo Thesaurus* is used. It also helps in word stemming.

The next step is to make the *Feature Vector (FV)* of the sentence $s_{ij}$ of text document $TD_i$ as

$$FV = \{F_{i11}, F_{i12}, F_{i13}, \ldots, F_{i1r}\} \quad (3)$$

Where $1 \leq i \leq n$, $1 \leq j \leq m$, and *r* depicts the total number of present features in the $s_{ij}$. *Feature Frequency Calculator (FFC)* calculates the frequency of each different feature occurred in FV. Frequency represents the number of occurrences of a feature in the sentence. So, each different feature is associated with its frequency in the form of a *Feature Frequency pair as* $<F_{ijk}, freq(F_{ijk})>$, where *i* is the text document number, *j* is the sentence number of the sentence, *k* is the feature number, *freq* () is a function to calculate the frequency of a feature, and $1 \leq k \leq r$.

The next step is to convert the high dimensional FV into low dimensional *Reduced Feature Vector (RFV)* to reduce the storage and execution time complexities. So, a counter loop is invoked to remove the duplicate or redundant entries of a feature. Therefore, only one instance of each different feature occurred is stored in RFV. It highly reduces the FV dimension and increases the efficiency of the system with good performance. The complete sentence level processing is shown in the Fig. 2.

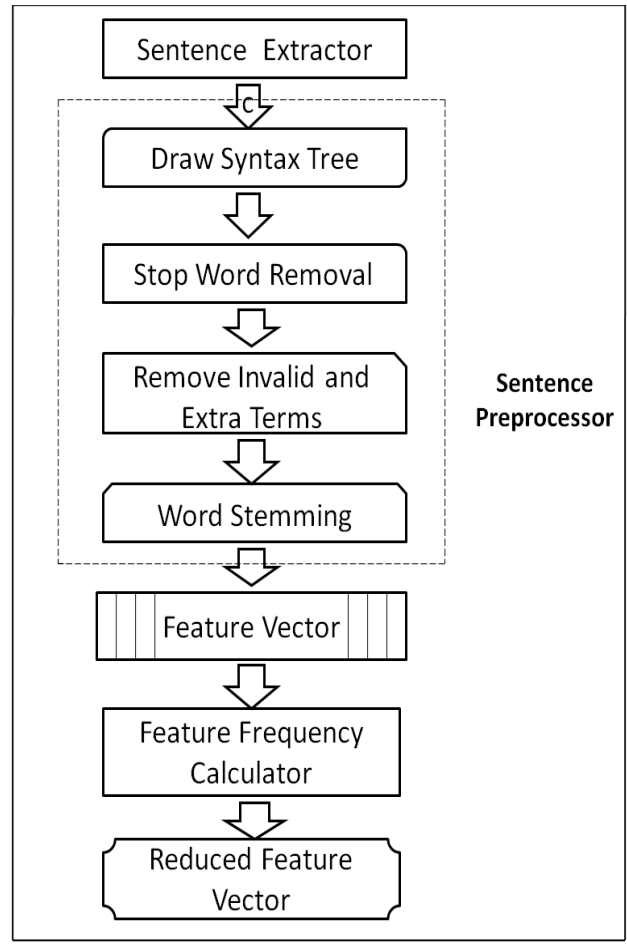

Figure 2. Text Document Preparation at Sentence Level.

Such sentence level preprocessing for the TDP is performed for each sentence of each document and then progressing toward the integrated corpora.

#### b) At Document Level

This step accepts the resultant RFV which is sent for document level pre-processing. Firstly, a counter loop is invoked to match the similar features in each of the $RFV_j$ with every other $RFV_q$ of a $TD_i$ where $1 \leq j, q \leq m, j \neq q$ and $1 \leq i \leq n$. *Refined Feature Vector Calculator (RFVC)* updates each feature's frequency for those features which are present two or more times in more than two sentences. These updates are done by adding up their frequencies in terms of combined calculated frequency of that feature only. In this way, it updates the count of each different occurred feature with more than one occurrence. The features that have occurred only once in the document will not update their frequency.

The next step is that all RFVs of a $TD_i$ are integrated into one as

$$IRFV = Integrat\ (RFV_1, RFV_2, ..., RFV_m) \quad (4)$$

Where *Integrat ()* is a function to combine all RFVs.

Now each $RFV_j$ is compared with every other $RFV_q$ where $1 \leq j, q \leq m$, and $j \neq q$. In this way, each feature of $RFV_j$ is compared with the every other feature of the $RFV_q$ and thereby, the duplicate and redundant features are removed. The complete procedure on the document level is shown in Fig. 3.

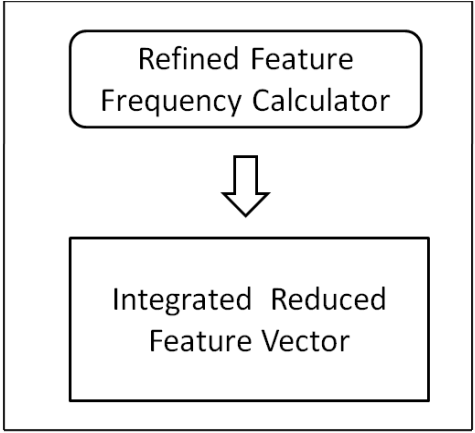

Figure 3. Text Document Preparation at Document Level.

*c) At Integrated Corpora*

In integrated corpora, all the IRFVs of *n* documents are integrated into one. This step is used to calculate and update the final frequency of each different feature occurred in the corpora. Firstly, it removes duplicate and redundant feature entries of IRFV, and then removes all ambiguous words with the help of the pseudo thesaurus. In such a way, *Integrated Corpora Feature Vector (ICFV)* is generated. A *Threshold Value (TV)* is defined for ICFV. TV cuts off those features whose *total frequency is less than TV*. Finally, it reduces the dimension of ICFV.

Therefore, *Integrated Corpora Feature Vector (ICFV)* is constructed as

$$ICFV = \{F_1, F_2, ..., F_n\} \quad (5)$$

Where $F_1, F_2, ..., F_n$ represent the features.

Such features with their associated frequencies show the statistical similarity between the text documents. Each feature is counted for each document and represented as the feature and its frequency as follows.

$$TD_1 = \{<F_1, f(F_1)>, <F_2, f(F_2)>, ..., <F_n, f(F_n)>\} \quad (6)$$

Where $f(F_i)$ is the function to calculate the frequency of feature $F_i$ in the text document.

The next step is to generate the matrix of TD and ICFV in the given form of table 1. In this matrix, the 0 represents the absence of that feature in TD and the numerical value represents the total number of occurrences of the feature in the TD.

TABLE I. TD ICFV MATRIX

| Text Document | Feature | | |
|---|---|---|---|
| | *F1* | *F2* | *F3* |
| TD1 | 1 | 0 | 1 |
| TD2 | 1 | 3 | 0 |
| TD3 | 0 | 2 | 1 |
| TD4 | 4 | 0 | 0 |

*2) Pseudo Thesaurus*

The Pseudo Thesaurus is a predefined English Vocabulary Set which is used to check the invalid words or to remove extra words from a sentence while processing the sentence in the TDTP. It is also used for word stemming so that the exact word can be obtained. For example, consider three different words for the word *research* - researching, researcher and researches. When the word stemming is performed, *research* is the final resulting word with the feature frequency counted as 3.

*3) Class Feeder (CF)*

Text Classification is the process of assigning the name of the class to a particular input, to which it belongs. The classes, from which the classification procedure can choose, can be described in many ways. So classification is considered as the essential part of many problems solving tasks or recognition tasks. Before classification can be done, the desired number of classes must be defined.

*4) Fuzzy Fetaure Category Similarity Analyzer (FFCSA)*

In FFCSA, firstly the *Feature Pattern (FP)* is made in the form of the membership degree of each feature with respect to every class. Consider text document set *TD* of *n* text documents as per given in equation 1, together with the ICFV *F* of *y* features $f_1, f_2, ... f_y$ and *e* classes $c_1, c_2, ..., c_e$. To construct the FP for each feature $f_k$ in F, its FP $fp_i$ is defined, by

$$fp_i = <fp_{i1}, fp_{i2}, fp_{i3}, ..., fp_{ie}> \quad (7)$$
$$= <\mu(f_i, c_1), \mu(f_i, c_2), \mu(f_i, c_3), ..., \mu(f_i, c_e)>$$

Where,

$f_i$ represents the number of occurrences of $f_i$ in the text document $TD_g$ where $1 \leq g \leq n$.

$\mu(f_i, c_e)$ is defined as the sum of product of the feature value of $f_i$ present in *n* text documents TD, w. r. t. a column vector and the 1 or 0 as the presence or absence of that feature in class $c_e$ / Sum of the feature value for the class $c_e$ only, where $1 \leq j \leq e$. So,

$$\mu(f_i, c_e) = \frac{\sum_g (TD_{gi}).b_i}{\sum_g (TD_{gi})} \quad (8)$$

$b_i$ is represented as
$$b_i = 1,\ \text{if}\ document\ \epsilon\ class\ c_e \quad (9)$$
$$= 0,\ otherwise$$

Each text document *TD* belongs to only one class *c*. In this way, each class can belong to one or more text documents. A set of n documents and their related categories or classes are represented as an ordered pair as shown

$$TD = \{<TD_1, C(TD_1)>, <TD_2, C(TD_2)>, ..., <TD_m, C(TD_m)>\} \quad (10)$$

Where the class of the text document $TD_i$: $C(TD_i) \in C$, $C(TD)$ is a categorization function whose domain is *TD* and range is *C*.

Each document belongs to one of the classes in the *C (TD)*. The resulted text documents that have many features are stored with their relevant classes as shown in the table 2. They are in the form of <*doc no, number of occurrences of each feature, class no*>.

TABLE II. TEXT DOCUMENT FEATURE VECTOR CLASS MATRIX

| Text Document | Feature | | | Class |
|---|---|---|---|---|
| | *F1* | *F2* | *F3* | |
| TD1 | 1 | 0 | 1 | C1 |
| TD2 | 1 | 3 | 0 | C2 |
| TD3 | 0 | 2 | 1 | C2 |
| TD4 | 4 | 0 | 0 | C3 |

In such a way, the relation between a feature and a class is made. Sometimes, it is quite possible that one document belongs to two or more classes that concern has to be considered by making the more presences of the text document in the table with the cost of increased complexity, so it is required to check each feature's distribution among them.

*B. Support Vector Machine Classifier (SVMC)*

The next step is to use the *Support Vector Machine Classifier (SVMC)*. SVMC is a popular and better method than other methods for text categorization. It is a kernel method which finds the maximum margin hyper plane in the feature space paradigm separating the data of training patterns into two groups like Boolean Logic 1 and 0. If any training pattern is not correctly classified by the hyper plane, then the concept of slack measure is used to get rid out of it.

Using this idea, SVMC can only separate apart two classes for $h = +1$ and $h = -1$. For e classes, one SVM for each class is created. For the SVM of class $c_l$, $1 \leq l \leq e$, the training patterns of class $c_l$ are for the $h = +1$ and of other classes are $h = -1$. The SVMC is then the aggregation of these SVMs.

SVM provides good results then KNN method because it directly divide the training data according to the hyper planes.

*C. Text Classification Phase*

To check the predictive accuracy of the SVMC, new and unknown text document is used, which is independent of the training text documents and is not used to construct the SVMC. The accuracy of this document is compared with the learned SVMC's class. If the accuracy of the SVMC is acceptable and good, then it can be used further to classify the future unseen text documents for which the class label is not known. Therefore, they can be categorized into the appropriate and a suitable category group.

IV. CONCLUSION AND FUTURE SCOPE

Text classification is expected to play an important role in future search services or in the text categorization. It is an essential matter to focus on the main subjects and significant content. It is becoming important to have the computational methods that automatically classify available text documents to obtain the categorized information or groups with greater speed and fidelity for the content matter of the texts.

As the proposed FSCMM model is made for text document categorization, it works well with high efficiency and effectiveness. Although this model and methodology seem very complex, yet it achieves the task of text categorization with high performance, and good accuracy and prediction. Feature Reduction is performed on the sentence, document and integrated corpora levels to highly reduce feature vector dimension. Such reduction improves the system performance greatly in terms of space and time. Result shows that the feature reduction reduces the space complexity by 20%.

Fuzzy similarity measure and methodology are used to make the matching connections among text documents, feature vectors and pre-defined classes. It provides the mathematical framework for finding out the membership degrees as feature frequency.

This model shows better results than other text categorization techniques. SVM classifier gives better results than KNN method. The system performance does not show high information gain and prediction results when KNN is used because it produces noise sensitive contents.

In the future, such model can be further extended to include the non-segmented text documents. It can also be extended to categorize the images, audio and video - related data.

AUTHORS PROFILE

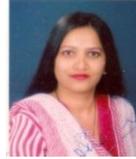

Shalini Puri is pursuing M.Tech in Computer Science at Birla Institute of Technology, Mesra, Ranchi, India. She is currently working as an Assistant Professor in a reputed engineering college in India. Her research areas include Artificial Intelligence, Data Mining, Soft Computing, Graph Theory, and Software Engineering.